\documentclass[sn-nature]{sn-jnl}

\pdfoutput=1
\usepackage{graphicx}%
\usepackage{multirow}%
\usepackage{amsmath,amssymb,amsfonts}%
\usepackage{amsthm}%
\usepackage{mathrsfs}%
\usepackage[title]{appendix}%
\usepackage{xcolor}%
\usepackage{textcomp}%
\usepackage{manyfoot}%
\usepackage{booktabs}%
\usepackage{algorithm}%
\usepackage{algorithmicx}%
\usepackage{algpseudocode}%
\usepackage{listings}%

\theoremstyle{thmstyleone}%
\theoremstyle{thmstyletwo}%

\theoremstyle{thmstylethree}%

\begin{document}

\title[Article Title]{Tracing Rayleigh-Taylor instability from measured periodic modulation in laser driven proton beams}


\author[1,2]{\fnm{Z.} \sur{Liu}}

\author[2,3]{\fnm{M.K.} \sur{Zhao}}

\author[2,3]{\fnm{P.L.} \sur{Bai}}

\author[2]{\fnm{X.J.} \sur{Yang}}

\author[2]{\fnm{R.} \sur{Qi}}

\author[2]{\fnm{Y.} \sur{Xu}}

\author[2]{\fnm{J.W.} \sur{Wang}}\email{wangjw@siom.ac.cn}

\author[2,3]{\fnm{Y.X.} \sur{Leng}}\email{lengyuxin@siom.ac.cn}

\author[2,3]{\fnm{J.H.} \sur{Bin}}\email{Jianhui.bin@siom.ac.cn}

\author[2,3,4]{\fnm{R.X.} \sur{Li}}

\affil[1]{\orgdiv{School of Microelectronics}, \orgname{Shanghai University}, \orgaddress{\city{Shanghai}, \postcode{201800}, \country{China}}}

\affil[2]{\orgdiv{State Key Laboratory of High Field Laser Physics}, \orgname{Shanghai Institude of Optics and Fine Mechanics, Chinese Academy of Sciences}, \orgaddress{ \city{Shanghai}, \postcode{201800}, \country{China}}}

\affil[3]{\orgdiv{Center of Materials Science and Optoelectronics Engineering}, \orgname{University of Chinese Academy of Sciences}, \orgaddress{\city{Beijing}, \postcode{101408}, \country{China}}}

\affil[4]{\orgdiv{School of School of Physical Science and Technology}, \orgname{ShanghaiTech University}, \orgaddress{\city{Shanghai}, \postcode{201210}, \country{China}}}


\abstract{Rayleigh-Taylor (RT) instability occurs in a variety of scenario as a consequence of fluids of different densities pushing against the density gradient. For example, it is expected to occur in the ion acceleration of solid density targets driven by high intensity lasers and is crucial for the acceleration process. Yet, it is essential to understand the dynamics of the RT instability, a typical way to measure this phenomenon requires sophisticated diagnostics such as streak X ray radiography. Here, we report on experimental observation on periodic modulation in the energy spectrum of laser accelerated proton beams. Interestingly, theoretical model and two-dimensional particle-in-cell simulations, in good agreement with the experimental finding, indicated that such modulation is associated with periodic modulated electron density induced by transverse Rayleigh-Taylor-like instability. Furthermore, the correlation between the RT instability and the ion acceleration provides an interpretation to trace the development of the RT instability from the modulated proton spectrum.
Our results thus suggest a possible tool to diagnose the evolution of the RT instability, and may have implications for further understanding for the accelerating mechanisms as well as optimization strategies for laser driven ion acceleration.}

\maketitle


As a consequence of fluids of different densities pushing against the density gradient,Rayleigh-Taylor (RT) instability occurs every scale from high energy density plasma \cite{cole1982measurement, isobe2005filamentary}, inertial confinement fusion (ICF)\cite{hurricane2014fuel, hurricane2016inertially}  to supernova explosion \cite{kuranz2018high, grefenstette2014asymmetries} and plays an important role in a wide range of physical process. For example, it is recognized to be one of the most significant limitation for ICF \cite{betti1998growth,weber2016first,sauppe2020demonstration}. Also, it occurs in the evolution from supernova explosion to remnant and strongly affect the formation of the structure \cite{kuranz2018high}. Understanding this phenomenon is crucial, so far the typical way to diagnose this phenomenon requires sophisticated diagnostics such as streak X-ray radiography \cite{cole1982measurement, rigon2021micron}, posting technical challenge to the experimental methods.

In recent years, large efforts have been devoted to explore the effort of RT instability on the process of laser driven ion acceleration \cite{palmer2012rayleigh, sgattoni2015laser, wan2018physical}. Laser driven ion acceleration has attracted much attention in the past two decades. Ideally, it offers an cost-effective solution for a compact accelerator and can be used in a wide range of applications, such as biomedical radiotherapy \cite{bin2022new,kroll2022tumour}, proton radiography \cite{dromey2016picosecond}, fast ignition fusion\cite{roth2001fast}, material science\cite{barberio2018laser}. When a high intensity laser pulse irradiates a solid density target, it can efficiently accelerate ions to high energies to 10s of MeV via various mechanism \cite{macchi2013ion, daido2012review, borghesi2006fast,schreiber2016invited}. Meanwhile, it can trigger transverse instability, such as RT instability and significant disturb the acceleration process. Simulation works have already shown that the RT instability can induce density ripples of target and lead to abortion of acceleration phase afterwards \cite{wan2016physical,wan2020effects}.

In this work, we report on experimental observation on periodic modulation in the energy spectrum of laser accelerated proton beams. Interestingly, theoretical model and two-dimensional particle-in-cell simulations, in good agreement with the experimental finding, indicated that such modulation is associated with periodic modulated electron density induced by transverse Rayleigh-Taylor-like instability. Furthermore, the correlation between the RT instability and the ion acceleration provides an interpretation to trace the development of the RT instability from the modulated proton spectrum.

\begin{figure}[htbp]
	\centering
	\includegraphics[width=1\linewidth]{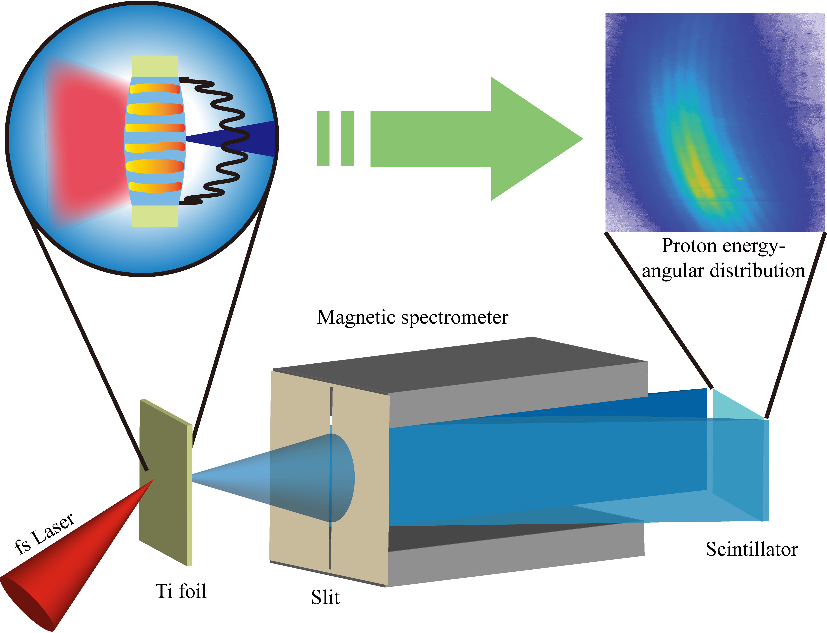}
	\caption{Experimental setup. An intense, femtosecond laser pulse (red) is focused into a Ti foil (green).  Accelerated proton bunch is deflected and characterized using a magnetic spectrometer \cite{bin2013small}. A raw image of the recorded proton energy-angular distribution containing periodic modulation is shown at scintillator position. }\label{fig1}
\end{figure}

The experimental setup is shown in Fig. \ref{fig1}. The experiments were performed with a  200 TW Ti:Sapphire laser system at SIOM in China \cite{xu2016stable,wu2020performance}. The temporal contrast of the 200 TW laser system was recently improved in both picosecond and nanosecond scale by using novel nonlinear temporal filter \cite{yu2018high,hu2021performance} and multi-pass amplifier configuration\cite{wang2022suppressing}, respectively. The 30 femtosecond laser pulse, was focused onto the target by a f/2 off-axis parabolic mirror, resulting in a peak intensity of $5\times10^{20}$ ${\rm W/cm^{2}}$ which corresponds to the normalized vector potential amplitude $a_0=15$. Titanium foils of thickness 1 $ \mu m $ have been irradiated at a 60 degree angle of incidence for varying laser pulse durations. A magnetic spectrometer with a long vertical entrance slit was used to detect protons along the target normal, enables angularly-resolved energy distribution of the accelerated protons (see Methods).

\begin{figure}[htbp]
	\centering	\includegraphics[width=1\linewidth]{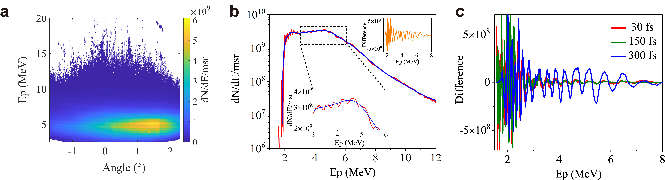}
	\caption{\textbf{Observed periodic modulation in the energy distribution of the accelerated protons.} \textbf{a}, Experimentally processed result of the data presented in Fig. \ref{fig1}. \textbf{b}, Measured proton spectrum along target normal (0 degree) is extracted from Fig. \ref{fig2}a (red curve), the blue curve shows the average trend of measured proton spectrum. The modulation feature is clearly visible by subtracting the average trend from the measured proton spectrum (inset). \textbf{c}, The energy modulation feature for varying laser pulse duration.}\label{fig2}
\end{figure}

A raw image from the 1$ \mu m $ Ti foil is shown in the inset of Fig. \ref{fig1}, a periodic modulation feature is clearly visible from the recorded raw image. The processed proton energy distribution from the experimental measured data is shown in Fig. \ref{fig2}a, by transferring the recorded two dimensional spatial information from the raw image to the angularly resolved energy distribution. The periodic modulation is presented over the full observable angular range of 4 degrees. Next, when we extracted the one-dimensional (1D) proton spectrum along the target normal (red curve in Fig. \ref{fig2}b), the modulation feature appears as weakly periodic variation on top of the overall decaying spectrum. By separating the measured proton spectrum from the average trend (blue curve in Fig. \ref{fig2}b), the modulation feature is further visualized by the difference. (inset in Fig. \ref{fig2}b). Note that, the weak modulation is hardly distinguishable from noise level if only a 1D proton spectrum is taken, for example, when a typical Thomson parabola spectrometer with a tiny entrance pinhole is applied. In our case, the periodic modulation is evident from our recorded two dimensional (2D) energy distribution of the protons.

Moreover, the modulation feature is observed for varying durations of the incident laser pulse, from initially 30 fs to 300 fs, with our given laser energy input of 4 J, a slight dependence of the amplitude related to the periodic variation is also observed for varying laser durations (Fig. \ref{fig2}c).

Spectral modulation in laser driven ion beams has been reported in previous experiments\cite{clark2000energetic, hegelich2002mev, ter2004spectral, ter2006quasimonoenergetic, fukuda2009energy, henig2009radiation, palmer2011monoenergetic, bin2015ion, kar2012ion, steinke2013stable, schnurer2013beat, noaman2018periodic}. Most observed modulations are presented as a few distinct peaks in the spectrum, and were associated to some perturbations on the accelerating field in the longitudinal dimension due to either multi-species effect in plasma expansion\cite{clark2000energetic, hegelich2002mev, ter2004spectral, ter2006quasimonoenergetic} or competition of multiple acceleration mechanisms\cite{fukuda2009energy, henig2009radiation, palmer2011monoenergetic, bin2015ion, kar2012ion, steinke2013stable}. Only a limited number of experimental results have been reported on spectral modulation with relatively high frequency similar to our results\cite{schnurer2013beat, noaman2018periodic}, interestingly, both were measured by a spectrometer with relatively larger acceptance angle. The high frequency spectral modulation is attributed to high frequency perturbations on the longitudinal accelerating field. 

For simplicity, in all the afore-mentioned results, laser driven ion acceleration is treated as purely 1D accelerating process, however, the actual accelerating is a complex, three-dimensional (3D) process. Translating the 1D concept to the actual multi-dimensional scenario,  one would expect modulated proton spectrum from perturbed accelerating field in the transverse dimension, similar to the longitudinal case.

\begin{figure}[htbp]
	\includegraphics[width=1\linewidth]{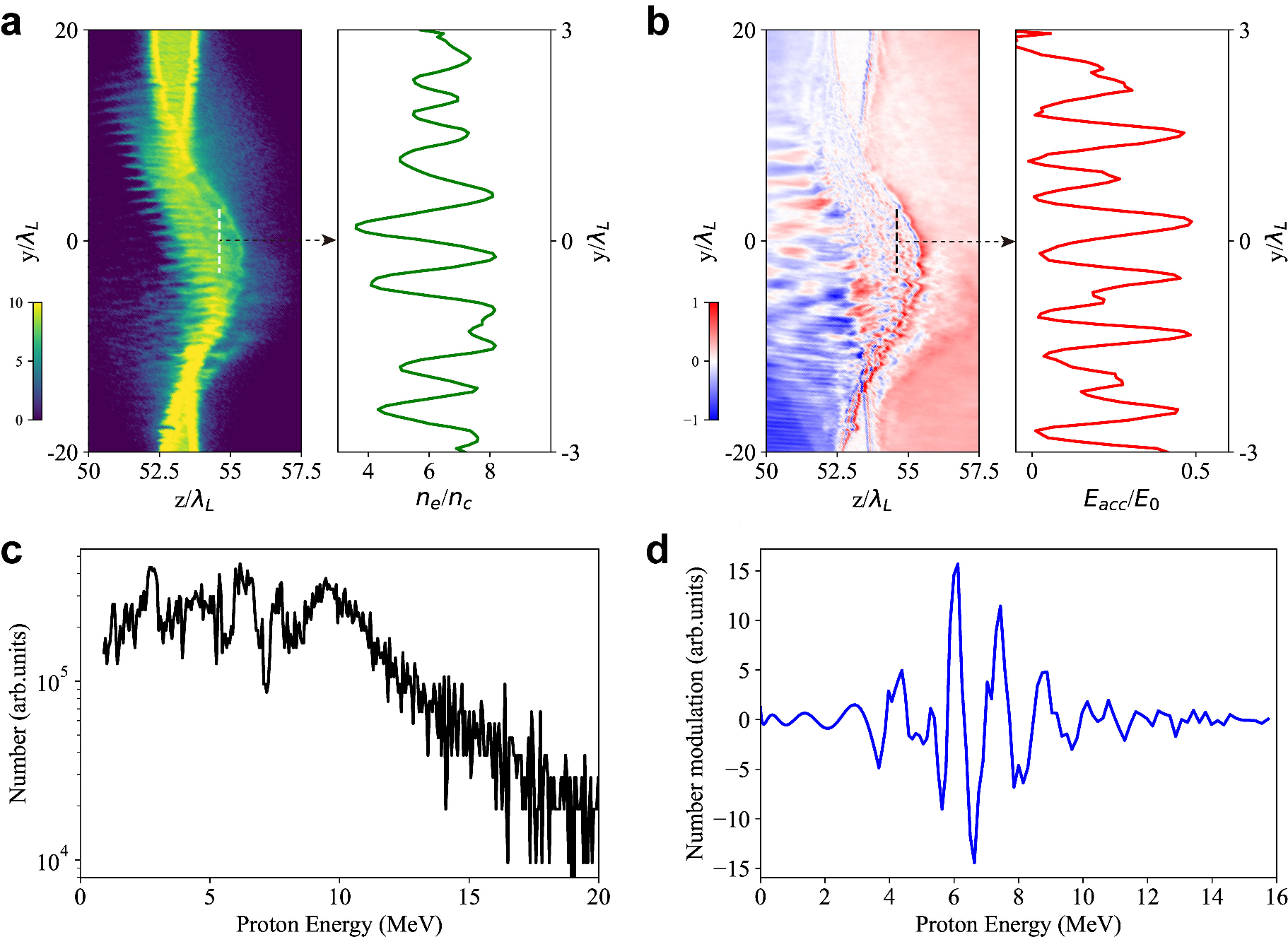}
	\centering
	\caption{{\textbf{Simulation results.}}  \footnotesize{\textbf{a}, Distribution of the electron density $n_e/n_c$ and its lineout distribution at $z$=54\rm{$\lambda_L$} (the white dashed line) at 350 fs after the pulse first reaches the target. \textbf{b}, Distribution of the electrostatic field ($E_L/(m_e\omega_L c/e)$) along the normal direction of the target and its lineout distribution at $z$=54\rm{$\lambda_L$} (the black dashed line) at 350 fs after the pulse first reaches the target. \textbf{c}, The proton spectrum and \textbf{d}, the corresponding number modulation at 700 fs after the pulse first reaches the target.}}\label{fig3}
\end{figure}

To validate our hypothesis, we carried out 2D particle-in-cell (PIC) simulations using the code LAPINE~\cite{bib23} (see Methods for detailed information). The simulation results are summarized in Fig. \ref{fig3}. Fig. \ref{fig3}a shows the density distribution of the electrons at 350fs after the pulse first reaches the target. The electrons are pushed forward by the laser while the density is obviously modulated transversely. The lineout distribution at $z$=54\rm{$\lambda_L$} indicates that the characteristic wavelength of the transverse density modulation is about $0.8\rm{\lambda_L}$. Those multiple density peaks and the characteristic wavelength (close to incident light wavelength) are typical features developed by transverse instabilities such as RT instability, similar to the previous observation \cite{palmer2012rayleigh}. The electrostatic field along the normal direction of the target is shown in Fig. \ref{fig3}b. At the rear side of the target, one can see two areas separated by a curved boundary. While the electrostatic field is more uniform on the right side of the boundary, the field is much more modulated on the left side. The modulation wavelength of the field is the same as the transverse electron density. Accelerated by such a modulated field, the spectrum of the proton (Fig. \ref{fig3}c) demonstrates a modulation feature. The modulation extracted from the spectrum is presented in Fig. \ref{fig3}d. As expected, a modulated transverse accelerating field lead to a modulated proton spectrum. One can see that the spectrum during the range  of 2-10 MeV is mostly modulated with a modulation period 1$\sim$2~MeV, in agree with our experimental data. In addition, the PIC simulation reveals that the transverse perturbation on the accelerating field is induced by transversely modulated electron density, a consequence from transverse RT instability developed during the laser plasma interaction. 

\begin{figure}[htbp]
	\centering
	\includegraphics[width=1\linewidth]{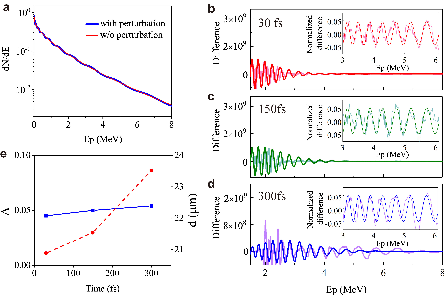}
	\caption{\textbf{Calculated proton energy distribution and estimated correlation on RT instability.} \textbf{a} Calculated proton energy distribution with (red) and w/o (blue) transverse electron density perturbation based on a plasma expansion model (Methods). Similar to Fig. \ref{fig2}c, the difference separated from the average trend, which represents the modulations on the proton spectrum, is plotted in \textbf{b-d} and compared with the best fitted energy distributions estimated from the analytical model for varying laser durations of 30 fs (\textbf{b}), 150 fs (\textbf{c}), and 300 fs \textbf{d}, respectively. The fitting parameters, $A$ and $d$, which represents the modulated amplitude induced by the RT instability and the spatial range of the RT instability, are plotted in \textbf{e}. In fact, $A$ is directly extracted from the difference normalized to average trend of the proton spectrum (insets of Fig. \ref{fig4}b-d). }\label{fig4}
\end{figure}

A simple analytical model is used to further explain our results (Methods). Starting from given electron density distributions, we calculated the corresponding proton energy distributions based on plasma expansion model and plotted in Fig. \ref{fig4}a. By inserting a perturbation term into the transverse electron density distribution, a modulation feature appears on the initially smooth, exponentially decaying spectrum, in well agreement with our experimental and simulation results. Indeed, it confirms our above interpretations, due to the multi-dimension nature of the laser driven ion acceleration, the modulated proton spectrum can be caused by a RT instability induced transversely modulated electron density, which is the main cause under our experimental condition, as supported by the simulations (Fig. \ref{fig3}). 

Further, the correlation between the measured proton spectrum and the RT instability can be also estimated based on the simple model. From the best fit results of the experimental data (Fig. \ref{fig4}b-d), we obtain the fitting parameters, the modulated amplitude of transverse electron density $A$ and the size of the interaction range $d$ where the RT instability developed, and plotted in Fig. \ref{fig4}e. Both values reflect the behave of RT instability under different experimental parameters, and provide a potential tool to trace the evolution of the RT instability. For instance, Fig. \ref{fig4}e indicates a weakly dependence of $A$ on the laser duration, while an enlarged spatial range $d$ for increasing laser duration. Those results highlight a potentially simple way to trace the evolution of the RT instability and motivate our future investigations. 

\newpage
\noindent \textbf{Methods}

\noindent \textbf{Experimental setup} 
The experiments was performed with a 200 TW Ti-sapphire laser system at Shanghai Institute of Optics and Fine mechanics (SIOM) in China. The system delivers pulses with a duration of 30 fs full-width at half-maximum (FWHM) centered at 800 nm wavelength. A 60$^{o}$ f/2 off-axis parabolic (OAP) mirror was used to focus the laser pulses to a measured FWHM focal spot diameter of 4 micrometers. A peak intensity of $5\times10^{20}$ $\rm W/cm^{2}$ is achieved with 4 Joules of laser energy, corresponding to a normalized vector potential amplitude $a_0=15$.

In our experiments, freestanding titanium foils of 1 $ \mu m $ thickness were irradiated at a 60 degree angle of incidence. A magnetic spectrometer with a gap of 14 cm and a long vertical slit of 400 $ \mu m $ width is placed after the target normal. A scintillator (EJ-200, Eljen Technology) located behind the magnets was used to measure the angularly-resolved energy distribution of protons over an angular range of  4 degree. A layer of $ 32 \mu m $ of Al foil was added in front of the scintillator to block laser light and heavy ions. Protons with energies beyond 1.7 MeV are recorded. 

\noindent \textbf{Particle-in-cell simulations} 
A \emph{p}-polarized laser pulse is obliquely incident on a solid target with an incident angle $\alpha=60^\circ$. The envelope of the pulse is shaped with a function of $\sin^2(\pi t/T_0)~(0\leq t\leq T_0)$, with the pulse full duration $T_0=100 T_L$. $T_L=\lambda_L/c$ is the laser period, $\lambda_L=0.8~\mu {\rm m}$ is the laser wavelength and $c$ is the light speed in the vacuum. The normalized amplitude of the laser field $E_L$ is $a_0 = eE_L/m_e\omega_L c = 5.0$ (corresponding to an intensity of $5.4\times 10^{19}~\rm{W/cm}^2$), where $m_e$ is the electron mass and $e$ is the electron charge. The target is representative of a solid $2.5\lambda_L$ thick Ti slab, defined on the electron density of $7n_c$ with $n_c=1.1\times10^{21}\lambda_L(\mu {\rm m})/{\rm cm}^3$ the critical density. The size of the simulation box is $60\lambda_L(y)\times100\lambda_L(z)$ corresponding to grids $1200(y)\times2000(z)$, with 81 macro-particles per cell.

\noindent \textbf{Proton energy distribution calculation} 
Assuming an initial electron density $ n_{e}=n_{e0}\cdot\xi(x)\cdot\xi(y)$, with a longitudinal density distribution $\xi(x)$ and a transverse density distribution $\xi(y)$. $\xi(x)$ corresponds to a Boltzmann distribution $\xi(x)=\exp(-x/c_{s}t-1) $, where $ c_{s}=(k_{B}T_{e}/m_{i})^{1/2} $,  $k_{B}$ represents the Boltzmann constant, $ T_{e} $ is the electron temperature, and $ m_{i} $ is the ion mass. And the transverse density approximates to a bell shape $ \xi(y)=\exp(-4\ln2\cdot\frac{y^{2}}{D^{2}}) $, where D is the FWHM diameter of the transverse Gaussian profile, similar to \cite{bib19, bib22}. The perturbation is expressed by a sine function$A\sin(\frac{2\pi}{c_{1}}\cdot y) $, where $A$ represents the modulated amplitude induced by the RT instability and we chose the same value from PIC simulation $c_{1}=0.8\rm{\lambda_L}$ as the modulation cycle. The resulting electron density is thus written as

\begin{equation}
	n_{e}=n_{0}\cdot\xi(x)\cdot\xi(y)\cdot(1+A\sin(\frac{2\pi}{c_{1}}\cdot y)).
\end{equation}

Insert $n_{e}=n_{p}$ into the solution $dN/dE = n_{p0}/[\sqrt{2E\cdot k_{B}T_{e}}\cdot \omega_{pi}^2t^2]$ \cite{bib21}, one derives the proton energy distribution 
\begin{equation}
	\begin{split}
		dN/dE = & n_{0} \cdot \exp[-\sqrt{\frac{2E}{k_{B}T_{e}}}-4\ln2\cdot(\frac{y}{D})^{2}]\\&\times[1+A\sin(\frac{2\pi}{c_{1}}\cdot y)]/\sqrt{2E\cdot k_{B}T_{e}}.
	\end{split}
\end{equation}
Where $y=d-c_{2}E^{1/4} $ represents the position dependence on proton energy follow a parabolic transverse expansion \cite{bib26}, $c_{2} $ and $d$ are fitting parameters which denotes the coefficient of the dependence and the size of the interaction range where RT instability developed, respectively.

\section*{Acknowledgements}

This work was supported by the CAS Project for Young Scientists in Basic Research (YSBR060), the National Natural Science Foundation of China (61925507), Natural Science Foundation of China (11127901), CAS Youth Innovation Promotion Association.

\section*{Data availability}
The experimental simulation data that support the findings of this study are available from the corresponding author upon reasonable request.

\section*{Author contributions}

Data analysis was carried out by Z.L. and J.H.B. All the authors (Z.L., M.K. Z., P.L.B., X.J.Y., R.Q., Y.X., Y.X.L., J.H.B., and R.X.L.) contributed to the planning, implementation and execution of the experiments. J.W.W. performed the simulations. Z.L., and J.H.B., prepared the initial manuscript. All the authors (Z.L., M.K. Z., P.L.B., X.J.Y., R.Q., Y.X., J.W.W., Y.X.L., J.H.B., and R.X.L.) contributed revision to the final version.

\section*{Additional information}

\textbf{Competing financial interests}: The authors declare no competing financial interests. 


\bibliographystyle{sn-standardnature}
\bibliography{sn-bibliography}

\begin{thebibliography}{10}
\expandafter\ifx\csname url\endcsname\relax
  \def\url#1{\burl{#1}}\fi
\expandafter\ifx\csname urlprefix\endcsname\relax\def\urlprefix{URL }\fi
\providecommand{\bibinfo}[2]{#2}
\providecommand{\eprint}[2][]{\url{#2}}
\providecommand{\doi}[1]{\url{https://doi.org/#1}}
\bibcommenthead

\bibitem{cole1982measurement}
\bibinfo{author}{Cole, A.} \emph{et~al.}
\newblock \bibinfo{title}{Measurement of rayleigh--taylor instability in a
  laser-accelerated target}.
\newblock \emph{\bibinfo{journal}{Nature}} \textbf{\bibinfo{volume}{299}},
  \bibinfo{pages}{329--331} (\bibinfo{year}{1982}).

\bibitem{isobe2005filamentary}
\bibinfo{author}{Isobe, H.}, \bibinfo{author}{Miyagoshi, T.},
  \bibinfo{author}{Shibata, K.} \& \bibinfo{author}{Yokoyama, T.}
\newblock \bibinfo{title}{Filamentary structure on the sun from the magnetic
  rayleigh--taylor instability}.
\newblock \emph{\bibinfo{journal}{Nature}} \textbf{\bibinfo{volume}{434}},
  \bibinfo{pages}{478--481} (\bibinfo{year}{2005}).

\bibitem{hurricane2014fuel}
\bibinfo{author}{Hurricane, O.} \emph{et~al.}
\newblock \bibinfo{title}{Fuel gain exceeding unity in an inertially confined
  fusion implosion}.
\newblock \emph{\bibinfo{journal}{Nature}} \textbf{\bibinfo{volume}{506}},
  \bibinfo{pages}{343--348} (\bibinfo{year}{2014}).

\bibitem{hurricane2016inertially}
\bibinfo{author}{Hurricane, O.~A.} \emph{et~al.}
\newblock \bibinfo{title}{Inertially confined fusion plasmas dominated by
  alpha-particle self-heating}.
\newblock \emph{\bibinfo{journal}{Nature Physics}}
  \textbf{\bibinfo{volume}{12}}, \bibinfo{pages}{800--806}
  (\bibinfo{year}{2016}).

\bibitem{kuranz2018high}
\bibinfo{author}{Kuranz, C.~C.} \emph{et~al.}
\newblock \bibinfo{title}{How high energy fluxes may affect rayleigh--taylor
  instability growth in young supernova remnants}.
\newblock \emph{\bibinfo{journal}{Nature communications}}
  \textbf{\bibinfo{volume}{9}}, \bibinfo{pages}{1564} (\bibinfo{year}{2018}).

\bibitem{grefenstette2014asymmetries}
\bibinfo{author}{Grefenstette, B.} \emph{et~al.}
\newblock \bibinfo{title}{Asymmetries in core-collapse supernovae from maps of
  radioactive 44ti in cassiopeia a}.
\newblock \emph{\bibinfo{journal}{Nature}} \textbf{\bibinfo{volume}{506}},
  \bibinfo{pages}{339--342} (\bibinfo{year}{2014}).

\bibitem{betti1998growth}
\bibinfo{author}{Betti, R.}, \bibinfo{author}{Goncharov, V.},
  \bibinfo{author}{McCrory, R.} \& \bibinfo{author}{Verdon, C.}
\newblock \bibinfo{title}{Growth rates of the ablative rayleigh--taylor
  instability in inertial confinement fusion}.
\newblock \emph{\bibinfo{journal}{Physics of Plasmas}}
  \textbf{\bibinfo{volume}{5}}, \bibinfo{pages}{1446--1454}
  (\bibinfo{year}{1998}).

\bibitem{weber2016first}
\bibinfo{author}{Weber, C.} \emph{et~al.}
\newblock \bibinfo{title}{First measurements of fuel-ablator interface
  instability growth in inertial confinement fusion implosions on the national
  ignition facility}.
\newblock \emph{\bibinfo{journal}{Physical Review Letters}}
  \textbf{\bibinfo{volume}{117}}, \bibinfo{pages}{075002}
  (\bibinfo{year}{2016}).

\bibitem{sauppe2020demonstration}
\bibinfo{author}{Sauppe, J.~P.} \emph{et~al.}
\newblock \bibinfo{title}{Demonstration of scale-invariant rayleigh-taylor
  instability growth in laser-driven cylindrical implosion experiments}.
\newblock \emph{\bibinfo{journal}{Physical Review Letters}}
  \textbf{\bibinfo{volume}{124}}, \bibinfo{pages}{185003}
  (\bibinfo{year}{2020}).

\bibitem{rigon2021micron}
\bibinfo{author}{Rigon, G.} \emph{et~al.}
\newblock \bibinfo{title}{Micron-scale phenomena observed in a turbulent
  laser-produced plasma}.
\newblock \emph{\bibinfo{journal}{Nature communications}}
  \textbf{\bibinfo{volume}{12}}, \bibinfo{pages}{2679} (\bibinfo{year}{2021}).

\bibitem{palmer2012rayleigh}
\bibinfo{author}{Palmer, C.} \emph{et~al.}
\newblock \bibinfo{title}{Rayleigh-taylor instability of an ultrathin foil
  accelerated by the radiation pressure of an intense laser}.
\newblock \emph{\bibinfo{journal}{Physical review letters}}
  \textbf{\bibinfo{volume}{108}}, \bibinfo{pages}{225002}
  (\bibinfo{year}{2012}).

\bibitem{sgattoni2015laser}
\bibinfo{author}{Sgattoni, A.}, \bibinfo{author}{Sinigardi, S.},
  \bibinfo{author}{Fedeli, L.}, \bibinfo{author}{Pegoraro, F.} \&
  \bibinfo{author}{Macchi, A.}
\newblock \bibinfo{title}{Laser-driven rayleigh-taylor instability: Plasmonic
  effects and three-dimensional structures}.
\newblock \emph{\bibinfo{journal}{Physical Review E}}
  \textbf{\bibinfo{volume}{91}}, \bibinfo{pages}{013106}
  (\bibinfo{year}{2015}).

\bibitem{wan2018physical}
\bibinfo{author}{Wan, Y.} \emph{et~al.}
\newblock \bibinfo{title}{Physical mechanism of the electron-ion coupled
  transverse instability in laser pressure ion acceleration for different
  regimes}.
\newblock \emph{\bibinfo{journal}{Physical Review E}}
  \textbf{\bibinfo{volume}{98}}, \bibinfo{pages}{013202}
  (\bibinfo{year}{2018}).

\bibitem{bin2022new}
\bibinfo{author}{Bin, J.} \emph{et~al.}
\newblock \bibinfo{title}{A new platform for ultra-high dose rate
  radiobiological research using the bella pw laser proton beamline}.
\newblock \emph{\bibinfo{journal}{Scientific reports}}
  \textbf{\bibinfo{volume}{12}}, \bibinfo{pages}{1484} (\bibinfo{year}{2022}).

\bibitem{kroll2022tumour}
\bibinfo{author}{Kroll, F.} \emph{et~al.}
\newblock \bibinfo{title}{Tumour irradiation in mice with a laser-accelerated
  proton beam}.
\newblock \emph{\bibinfo{journal}{Nature Physics}}
  \textbf{\bibinfo{volume}{18}}, \bibinfo{pages}{316--322}
  (\bibinfo{year}{2022}).

\bibitem{dromey2016picosecond}
\bibinfo{author}{Dromey, B.} \emph{et~al.}
\newblock \bibinfo{title}{Picosecond metrology of laser-driven proton bursts}.
\newblock \emph{\bibinfo{journal}{Nature communications}}
  \textbf{\bibinfo{volume}{7}}, \bibinfo{pages}{1--6} (\bibinfo{year}{2016}).

\bibitem{roth2001fast}
\bibinfo{author}{Roth, M.} \emph{et~al.}
\newblock \bibinfo{title}{Fast ignition by intense laser-accelerated proton
  beams}.
\newblock \emph{\bibinfo{journal}{Physical review letters}}
  \textbf{\bibinfo{volume}{86}}, \bibinfo{pages}{436} (\bibinfo{year}{2001}).

\bibitem{barberio2018laser}
\bibinfo{author}{Barberio, M.} \emph{et~al.}
\newblock \bibinfo{title}{Laser-accelerated particle beams for stress testing
  of materials}.
\newblock \emph{\bibinfo{journal}{Nature communications}}
  \textbf{\bibinfo{volume}{9}}, \bibinfo{pages}{372} (\bibinfo{year}{2018}).

\bibitem{macchi2013ion}
\bibinfo{author}{Macchi, A.}, \bibinfo{author}{Borghesi, M.} \&
  \bibinfo{author}{Passoni, M.}
\newblock \bibinfo{title}{Ion acceleration by superintense laser-plasma
  interaction}.
\newblock \emph{\bibinfo{journal}{Reviews of Modern Physics}}
  \textbf{\bibinfo{volume}{85}}, \bibinfo{pages}{751} (\bibinfo{year}{2013}).

\bibitem{daido2012review}
\bibinfo{author}{Daido, H.}, \bibinfo{author}{Nishiuchi, M.} \&
  \bibinfo{author}{Pirozhkov, A.~S.}
\newblock \bibinfo{title}{Review of laser-driven ion sources and their
  applications}.
\newblock \emph{\bibinfo{journal}{Reports on progress in physics}}
  \textbf{\bibinfo{volume}{75}}, \bibinfo{pages}{056401}
  (\bibinfo{year}{2012}).

\bibitem{borghesi2006fast}
\bibinfo{author}{Borghesi, M.} \emph{et~al.}
\newblock \bibinfo{title}{Fast ion generation by high-intensity laser
  irradiation of solid targets and applications}.
\newblock \emph{\bibinfo{journal}{Fusion science and technology}}
  \textbf{\bibinfo{volume}{49}}, \bibinfo{pages}{412--439}
  (\bibinfo{year}{2006}).

\bibitem{schreiber2016invited}
\bibinfo{author}{Schreiber, J.}, \bibinfo{author}{Bolton, P.} \&
  \bibinfo{author}{Parodi, K.}
\newblock \bibinfo{title}{Invited review article:“hands-on” laser-driven
  ion acceleration: A primer for laser-driven source development and potential
  applications}.
\newblock \emph{\bibinfo{journal}{Review of Scientific Instruments}}
  \textbf{\bibinfo{volume}{87}} (\bibinfo{year}{2016}).

\bibitem{wan2016physical}
\bibinfo{author}{Wan, Y.} \emph{et~al.}
\newblock \bibinfo{title}{Physical mechanism of the transverse instability in
  radiation pressure ion acceleration}.
\newblock \emph{\bibinfo{journal}{Physical review letters}}
  \textbf{\bibinfo{volume}{117}}, \bibinfo{pages}{234801}
  (\bibinfo{year}{2016}).

\bibitem{wan2020effects}
\bibinfo{author}{Wan, Y.}, \bibinfo{author}{Andriyash, I.},
  \bibinfo{author}{Lu, W.}, \bibinfo{author}{Mori, W.} \&
  \bibinfo{author}{Malka, V.}
\newblock \bibinfo{title}{Effects of the transverse instability and wave
  breaking on the laser-driven thin foil acceleration}.
\newblock \emph{\bibinfo{journal}{Physical Review Letters}}
  \textbf{\bibinfo{volume}{125}}, \bibinfo{pages}{104801}
  (\bibinfo{year}{2020}).

\bibitem{xu2016stable}
\bibinfo{author}{Xu, Y.} \emph{et~al.}
\newblock \bibinfo{title}{A stable 200tw/1hz ti: sapphire laser for driving
  full coherent xfel}.
\newblock \emph{\bibinfo{journal}{Optics \& Laser Technology}}
  \textbf{\bibinfo{volume}{79}}, \bibinfo{pages}{141--145}
  (\bibinfo{year}{2016}).

\bibitem{wu2020performance}
\bibinfo{author}{Wu, F.} \emph{et~al.}
\newblock \bibinfo{title}{Performance improvement of a 200tw/1hz ti: sapphire
  laser for laser wakefield electron accelerator}.
\newblock \emph{\bibinfo{journal}{Optics \& Laser Technology}}
  \textbf{\bibinfo{volume}{131}}, \bibinfo{pages}{106453}
  (\bibinfo{year}{2020}).

\bibitem{yu2018high}
\bibinfo{author}{Yu, L.} \emph{et~al.}
\newblock \bibinfo{title}{High-contrast front end based on cascaded xpwg and
  femtosecond opa for 10-pw-level ti: sapphire laser}.
\newblock \emph{\bibinfo{journal}{Optics express}}
  \textbf{\bibinfo{volume}{26}}, \bibinfo{pages}{2625--2633}
  (\bibinfo{year}{2018}).

\bibitem{hu2021performance}
\bibinfo{author}{Hu, J.} \emph{et~al.}
\newblock \bibinfo{title}{Performance improvement of a nonlinear temporal
  filter by using cascaded femtosecond optical parametric amplification}.
\newblock \emph{\bibinfo{journal}{Optics Express}}
  \textbf{\bibinfo{volume}{29}}, \bibinfo{pages}{37443--37452}
  (\bibinfo{year}{2021}).

\bibitem{wang2022suppressing}
\bibinfo{author}{Wang, X.} \emph{et~al.}
\newblock \bibinfo{title}{Suppressing scattering-induced nanosecond pre-pulses
  in ti: sapphire multi-pass amplifiers}.
\newblock \emph{\bibinfo{journal}{Optics Letters}}
  \textbf{\bibinfo{volume}{47}}, \bibinfo{pages}{5164--5167}
  (\bibinfo{year}{2022}).

\bibitem{bin2013small}
\bibinfo{author}{Bin, J.} \emph{et~al.}
\newblock \bibinfo{title}{On the small divergence of laser-driven ion beams
  from nanometer thick foils}.
\newblock \emph{\bibinfo{journal}{Physics of Plasmas}}
  \textbf{\bibinfo{volume}{20}} (\bibinfo{year}{2013}).

\bibitem{clark2000energetic}
\bibinfo{author}{Clark, E.} \emph{et~al.}
\newblock \bibinfo{title}{Energetic heavy-ion and proton generation from
  ultraintense laser-plasma interactions with solids}.
\newblock \emph{\bibinfo{journal}{Physical Review Letters}}
  \textbf{\bibinfo{volume}{85}}, \bibinfo{pages}{1654} (\bibinfo{year}{2000}).

\bibitem{hegelich2002mev}
\bibinfo{author}{Hegelich, M.} \emph{et~al.}
\newblock \bibinfo{title}{Mev ion jets from short-pulse-laser interaction with
  thin foils}.
\newblock \emph{\bibinfo{journal}{Physical review letters}}
  \textbf{\bibinfo{volume}{89}}, \bibinfo{pages}{085002}
  (\bibinfo{year}{2002}).

\bibitem{ter2004spectral}
\bibinfo{author}{Ter-Avetisyan, S.} \emph{et~al.}
\newblock \bibinfo{title}{Spectral dips in ion emission emerging from
  ultrashort laser-driven plasmas}.
\newblock \emph{\bibinfo{journal}{Physical review letters}}
  \textbf{\bibinfo{volume}{93}}, \bibinfo{pages}{155006}
  (\bibinfo{year}{2004}).

\bibitem{ter2006quasimonoenergetic}
\bibinfo{author}{Ter-Avetisyan, S.} \emph{et~al.}
\newblock \bibinfo{title}{Quasimonoenergetic deuteron bursts produced by
  ultraintense laser pulses}.
\newblock \emph{\bibinfo{journal}{Physical Review Letters}}
  \textbf{\bibinfo{volume}{96}}, \bibinfo{pages}{145006}
  (\bibinfo{year}{2006}).

\bibitem{fukuda2009energy}
\bibinfo{author}{Fukuda, Y.} \emph{et~al.}
\newblock \bibinfo{title}{Energy increase in multi-mev ion acceleration in the
  interaction of a short pulse laser with a cluster-gas target}.
\newblock \emph{\bibinfo{journal}{Physical review letters}}
  \textbf{\bibinfo{volume}{103}}, \bibinfo{pages}{165002}
  (\bibinfo{year}{2009}).

\bibitem{henig2009radiation}
\bibinfo{author}{Henig, A.} \emph{et~al.}
\newblock \bibinfo{title}{Radiation-pressure acceleration of ion beams driven
  by circularly polarized laser pulses}.
\newblock \emph{\bibinfo{journal}{Physical Review Letters}}
  \textbf{\bibinfo{volume}{103}}, \bibinfo{pages}{245003}
  (\bibinfo{year}{2009}).

\bibitem{palmer2011monoenergetic}
\bibinfo{author}{Palmer, C.~A.} \emph{et~al.}
\newblock \bibinfo{title}{Monoenergetic proton beams accelerated by a radiation
  pressure driven shock}.
\newblock \emph{\bibinfo{journal}{Physical review letters}}
  \textbf{\bibinfo{volume}{106}}, \bibinfo{pages}{014801}
  (\bibinfo{year}{2011}).

\bibitem{bin2015ion}
\bibinfo{author}{Bin, J.} \emph{et~al.}
\newblock \bibinfo{title}{Ion acceleration using relativistic pulse shaping in
  near-critical-density plasmas}.
\newblock \emph{\bibinfo{journal}{Physical review letters}}
  \textbf{\bibinfo{volume}{115}}, \bibinfo{pages}{064801}
  (\bibinfo{year}{2015}).

\bibitem{kar2012ion}
\bibinfo{author}{Kar, S.} \emph{et~al.}
\newblock \bibinfo{title}{Ion acceleration in multispecies targets driven by
  intense laser radiation pressure}.
\newblock \emph{\bibinfo{journal}{Physical Review Letters}}
  \textbf{\bibinfo{volume}{109}}, \bibinfo{pages}{185006}
  (\bibinfo{year}{2012}).

\bibitem{steinke2013stable}
\bibinfo{author}{Steinke, S.} \emph{et~al.}
\newblock \bibinfo{title}{Stable laser-ion acceleration in the light sail
  regime}.
\newblock \emph{\bibinfo{journal}{Physical Review Special Topics—Accelerators
  and Beams}} \textbf{\bibinfo{volume}{16}}, \bibinfo{pages}{011303}
  (\bibinfo{year}{2013}).

\bibitem{schnurer2013beat}
\bibinfo{author}{Schn{\"u}rer, M.} \emph{et~al.}
\newblock \bibinfo{title}{The beat in laser-accelerated ion beams}.
\newblock \emph{\bibinfo{journal}{Physics of plasmas}}
  \textbf{\bibinfo{volume}{20}} (\bibinfo{year}{2013}).

\bibitem{noaman2018periodic}
\bibinfo{author}{Noaman-ul Haq, M.} \emph{et~al.}
\newblock \bibinfo{title}{Periodic spectral modulations of low-energy,
  low-charge-state carbon ions accelerated in an intense laser--solid
  interaction}.
\newblock \emph{\bibinfo{journal}{Physics of Plasmas}}
  \textbf{\bibinfo{volume}{25}} (\bibinfo{year}{2018}).

\bibitem{bib23}
\bibinfo{author}{Le~Graverand, M.-P.~H.} \emph{et~al.}
\newblock \bibinfo{title}{Assessment of specific mrna levels in cartilage
  regions in a lapine model of osteoarthritis}.
\newblock \emph{\bibinfo{journal}{Journal of orthopaedic research}}
  \textbf{\bibinfo{volume}{20}}, \bibinfo{pages}{535--544}
  (\bibinfo{year}{2002}).

\bibitem{bib19}
\bibinfo{author}{Schn{\"u}rer, M.} \emph{et~al.}
\newblock \bibinfo{title}{The beat in laser-accelerated ion beams}.
\newblock \emph{\bibinfo{journal}{Physics of plasmas}}
  \textbf{\bibinfo{volume}{20}}, \bibinfo{pages}{103102}
  (\bibinfo{year}{2013}).

\bibitem{bib22}
\bibinfo{author}{Bin, J.} \emph{et~al.}
\newblock \bibinfo{title}{Ultrasmall divergence of laser-driven ion beams from
  nanometer thick foils}.
\newblock \emph{\bibinfo{journal}{arXiv preprint arXiv:1303.2535}}
  (\bibinfo{year}{2013}).

\bibitem{bib21}
\bibinfo{author}{Mora, P.}
\newblock \bibinfo{title}{Plasma expansion into a vacuum}.
\newblock \emph{\bibinfo{journal}{Physical Review Letters}}
  \textbf{\bibinfo{volume}{90}}, \bibinfo{pages}{185002}
  (\bibinfo{year}{2003}).

\bibitem{bib26}
\bibinfo{author}{Brambrink, E.} \emph{et~al.}
\newblock \bibinfo{title}{Transverse characteristics of short-pulse
  laser-produced ion beams: A study of the acceleration dynamics}.
\newblock \emph{\bibinfo{journal}{Physical review letters}}
  \textbf{\bibinfo{volume}{96}}, \bibinfo{pages}{154801}
  (\bibinfo{year}{2006}).

\end{thebibliography}

\end{document}